\documentclass[showpacs,twocolumn,preprintnumbers,amsmath,amssymb,pra]{revtex4}

\usepackage{graphicx}
\usepackage{dcolumn}
\usepackage{bm}
\usepackage[latin1]{inputenc}
\usepackage{psfrag}
\usepackage{epic}
\usepackage{eepic}
\usepackage{bm}
\usepackage{mathrsfs}

\begin{document}
\newcommand{\trace}[1]{\mathrm{Tr}\, #1}
\newcommand{\vkt}[1]{{\boldsymbol{#1}}}
\newcommand{\matr}[1]{#1}
\newcommand{\matrg}[1]{#1}
\newcommand{\transp}{\mathrm{T}}
\newcommand{\forv}[1]{\langle #1\rangle}
\newcommand{\bra}[1]{\langle #1|}
\newcommand{\ket}[1]{|#1\rangle}
\newcommand{\braket}[2]{\langle #1|#2 \rangle}
\newcommand{\dd}{\ensuremath{\mathrm{d}}}
\newcommand{\imi}{\ensuremath{\mathrm{i}}}
\newcommand{\expe}{\ensuremath{\mathrm{e}}}
\newcommand{\Tr}{\mathrm{Tr}}
\newcommand{\expt}[1]{\left<#1\right>}
\newenvironment{smatrix}{\begin{bmatrix}}{\end{bmatrix}}

\renewcommand{\vec}[1]{\bm{#1}}
\setlength{\tabcolsep}{4ex}
\setlength{\unitlength}{1mm}

\title{Entanglement used to identify critical systems}

\author{S. O. Skrøvseth}
\email{stein.skrovseth@ntnu.no}
\author{K. Olaussen}
\email{kare.olaussen@ntnu.no}
\affiliation{%
Department of Physics,
Norwegian University of Science and Technology,
N-7491 Trondheim, Norway
}%

\date{10 March 2005}

\begin{abstract}
We promote use of the geometric entropy formula
derived by Holzhey {\em et. al.} from conformal field theory,
$S_\ell \sim ({c}/{3})\,\log\left(\sin{\pi\ell}/{N}\right)$, 
to identify critical
regions in zero temperature 1D quantum systems. The method is demonstrated
on a class of one-dimensional $XY$ and $XYZ$ spin-$\frac{1}{2}$ chains,
where the critical regions and their correponding central charges can be
reproduced with quite modest computational efforts.
\end{abstract}

\pacs{03.67.Mn, 75.10.Pq, 11.25.Hf}
\maketitle

\section{Introduction}

Continuous phase transitions \cite{Sachdev} are characterized by a diverging correlation length
and the emergence of scale invariance. If the system is governed by local
interactions (so that a conserved, symmetric and
traceless energy-momentum tensor can be defined at the critical points)
the scale symmetry is automatically enlarged to the full
conformal group \cite{Polyakov70}. In
two dimensions this group is infinite dimensional; this provides a lot
of information about the behaviour of critical systems in two
dimensions \cite{BPZ84a,BPZ84b}. One quantity which can be
computed exactly by use of conformal symmetry is the entanglement entropy,
i.e.\ the von Neuman entropy
\begin{equation}
  S_\mathcal A=-\trace\rho_\mathcal A\log\rho_{\mathcal A}
\end{equation}
of the reduced density matrix
$\rho_{\mathcal A} = \trace_{\mathcal B}
\vert\Omega\rangle\,\langle\Omega\vert$ for a
subsystem $\mathcal A$ of a system $\mathcal A+\mathcal B$ in the
ground state $\vert\Omega\rangle$. This entanglement entropy is, in
the case of pure states, a measure of the entanglement between
subsystems $\mathcal A$ and $\mathcal B$ \cite{Nielsen&Chuang}. In the case where
$\mathcal A$ is a segment of length $\ell$ of a
circle $\mathcal A+\mathcal B$ of circumference $N$ the result is
\begin{equation}
  s_\ell \equiv S_\ell - S_{N/2} =  
  \frac{c+\bar c}{6}\log\left(\sin\frac{\pi\ell}{N}\right),
  \label{Holzhey_formula}
\end{equation}
where $c$ ($\bar c$) is the holomorphic
(antiholomorphic) central charge. This result was first derived by
Holzhey {\em et.\ al.} \cite{Holzhey94} (denoted geometric entropy),
and has recently been extended
in several directions by Calabrese and Cardy \cite{Calabrese04}. The
explicitly known dependence of $s_\ell$ on the scaling variable
$\ell/N$ is a distinguishing feature of Eq.~(\ref{Holzhey_formula}).
\begin{figure}
  \begin{center}
    \psfrag{X}{$\ell/N$}
    \psfrag{Y}{\raisebox{1.5ex}{\hspace{0.5em}\large $\hat s_\ell$}}
    \psfrag{A}{$N=10$ (Critical)}
    \psfrag{B}{$N=10$ (Non-critical)}
    \psfrag{C}{$N=100$ (Non-critical)}
    \psfrag{D}{$s_\ell$ (with $c=\bar{c}=\frac{1}{2}$)}
    \includegraphics[angle=-90]{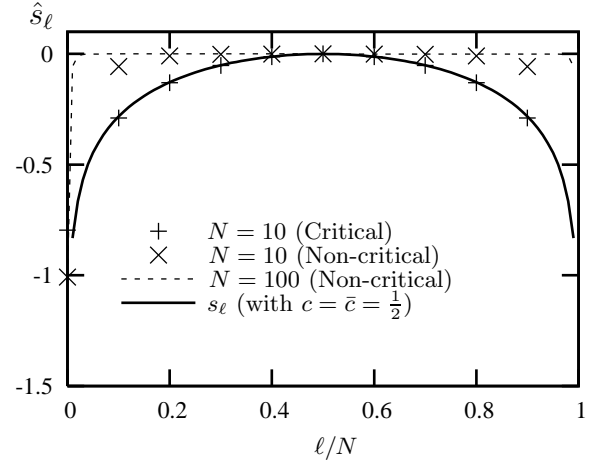}
  \end{center}
  \caption{Entanglement in
     the 1D quantum Ising model,
     Eq.~(\ref{H_XY}) with $\gamma=1$,
     for critical ($\lambda=1$) and noncritical ($\lambda=\frac{1}{2}$)
     parameter values, together with the
     predicted critical behavior.
     The critical $N=100$ line is indistinguishable
     from $s_\ell$ of
     Eq.~(\ref{Holzhey_formula}). 
   }
   \label{fig:Entanglement}
\end{figure}
\begin{figure}
   \begin{center}
     \psfrag{X1}{$\lambda$}
     \psfrag{Y1}{\large $\varepsilon$}
     \psfrag{A}{\hspace{-2.1em}$N\!=\!50$}
     \psfrag{B}{\hspace{-2.1em}$N\!=\!10$}
     \psfrag{-5}{\hspace{-1.48em}$10^{-5}$}
     \psfrag{-6}{\hspace{-1.48em}$10^{-6}$}
     \psfrag{-7}{\hspace{-1.48em}$10^{-7}$}
     \psfrag{-8}{\hspace{-1.48em}$10^{-8}$}
     \psfrag{-9}{\hspace{-1.48em}$10^{-9}$}
     \psfrag{-10}{\hspace{-1.0em}$10^{-10}$}
     \psfrag{-11}{\hspace{-1.0em}$10^{-11}$}
     \psfrag{-12}{\hspace{-1.0em}$10^{-12}$}
     \psfrag{-13}{\hspace{-1.0em}$10^{-13}$}
     \includegraphics[angle=-90]{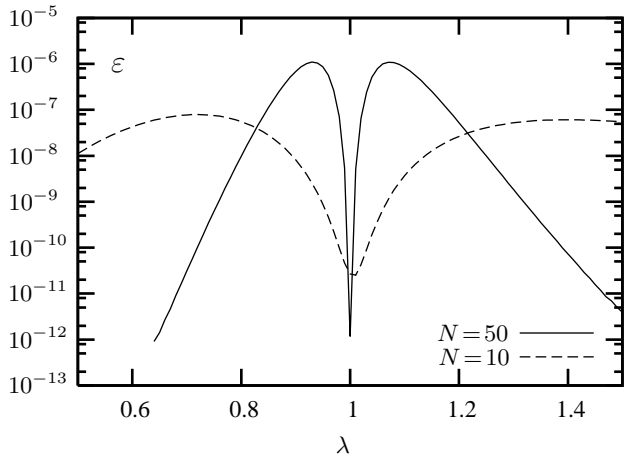}
   \end{center}
   \caption{
     The deviation $\varepsilon$ defined by Eq.~(\ref{Deviation}),
     calculated from Eq.~(\ref{H_XY}) with $\gamma=1$.
     The minimum identifies the critical point
     to about 0.5\% accuracy when $N=10$; see Fig.~\ref{fig:convergence}.
     The small value of $\varepsilon$ far away from the critical point
     occurs because a noncritical system can fit Eq.~(\ref{Holzhey_formula})
     with $c=\bar{c}=0$ (see Fig.~\ref{fig:Ising_c}).
   }
   \label{fig:Ising_err}
\end{figure}

\begin{figure}
   \begin{center}
     \psfrag{X}{$1/N$}
     \psfrag{Y1}{\raisebox{-1.5ex}{\hspace{-0.5em}$\hat\lambda_c$}}
     \psfrag{Y2}{\raisebox{-1.0ex}{\hspace{0.7em}$\varepsilon$}}
     \psfrag{1.02}{$1.02$}
     \psfrag{1.01}{$1.01$}
     \psfrag{1}{$1$}
     \psfrag{-8}{$10^{-8}$}
     \psfrag{-9}{$10^{-9}$}
     \psfrag{-10}{$10^{-10}$}
     \psfrag{-11}{$10^{-11}$}
     \psfrag{-12}{$10^{-12}$}
     \psfrag{-13}{$10^{-13}$}
     \includegraphics[angle=-90]{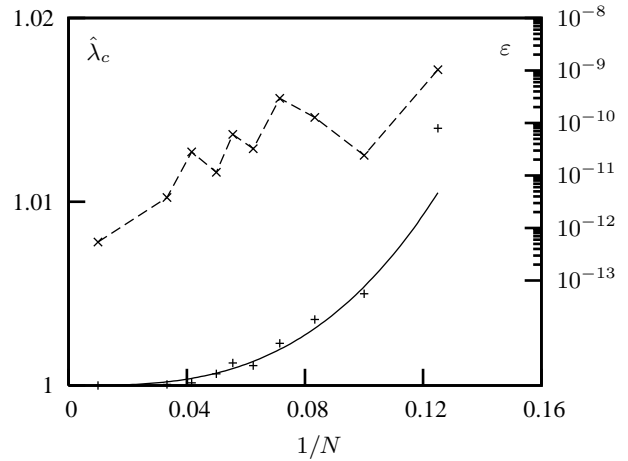}
   \end{center}
   \caption{The estimated critical point $\hat\lambda_c$ ($+$)
     and the corresponding deviation $\varepsilon$ ($\times$)
     as function of system size $N$. The fulldrawn line is the
     fit $\hat\lambda_c=1 + 5.37415\;N^{-3}$.
   }
   \label{fig:convergence}
\end{figure}

\begin{figure}[b]
   \begin{center}
     \psfrag{X2}{$\lambda$}
     \psfrag{Y2}{$\hat{c}$}
     \includegraphics[angle=-90]{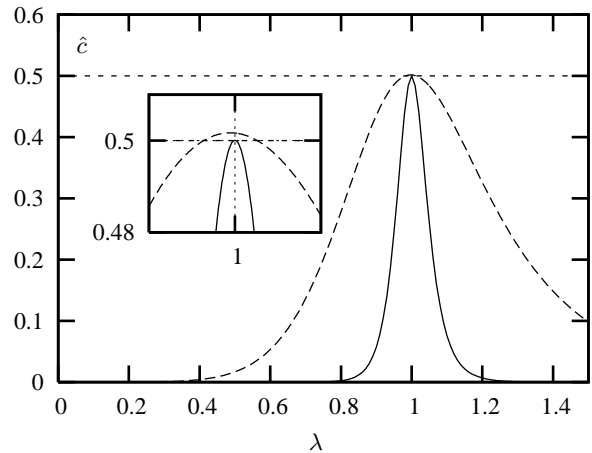}
   \end{center}
   \caption{The
     estimated central charge $\hat c$ for the 1D Ising model
     (cf.\ Fig.~\ref{fig:Ising_err}) obtained by fitting $c=\bar{c}$ so that
     $\varepsilon$ of Eq.~(\ref{Deviation}) is minimized. The dashed line
     is for $N=10$, the fulldrawn line for $N=50$.
     The inset shows details close to the critical point in 
     the 1 percent region. As can be seen the
     correct central charge is well reproduced.}
   \label{fig:Ising_c}
\end{figure}

The main purpose of this note is to point out how well Eq.~(\ref{Holzhey_formula})
is reproduced at critical parameters \cite{footnote1}
, even for very small systems. The
comparison of a numerical calculated $\ell$-dependence
with this formula is in our experience a very efficient method to identify critical points.
This is illustrated in Fig.~\ref{fig:Entanglement}.
As can be seen the entanglement of
noncritical systems saturates at finite $\ell$ (determined by the
correlation length), in contrast to critical ones which follow
Eq.~(\ref{Holzhey_formula}) closely. The 
system considered in Fig.~\ref{fig:Entanglement} is defined by the Hamiltonian
\begin{equation}
  H_{\mathrm{Ising}}=\sum_{n=0}^{N-1}\left(\sigma_{n}^x\sigma_{n+1}^x+\lambda\sigma_n^z\right),
  \label{QuantumIsing}
\end{equation}
where periodic boundary conditions are applied, i.e., $\sigma_{N}\equiv\sigma_0$.
We shall refer to it as the 1D quantum Ising model \cite{footnote2}.
Throughout this
paper logarithms are calculated base two.
As can be seen the critical point can be quite accurately determined from
surprisingly small systems. An additonal advantage is that the analysis
can be performed without varying the total system size $N$.
To further illustrate this point we have calculated the mean square deviation
\begin{equation}
  \varepsilon \equiv \frac{1}{M}\sum_{\ell} \left({\hat s}_\ell-s_\ell\right)^2,
  \label{Deviation}
\end{equation}
where $\ell$ runs over integers satisfying $0.2\,N < \ell < 0.8\,N$, $M$ is the number
of such integers, and ${\hat s}_\ell$ is the quantity
corresponding to Eq.~(\ref{Holzhey_formula})
calculated from the ground state of a finite system with $c=\bar{c}$ chosen to
minimize $\varepsilon$. Results for the quantum Ising model is shown in
Fig.~\ref{fig:Ising_err}, where the critical Ising point $\lambda=1$
is readily identified even for a system as small as $N=10$. Using the
minimal error as an estimate of the critical point we have plotted out
this estimate and the corresponding error in Fig.~\ref{fig:convergence}, 
which illustrates how the identification improves
with system size. In Fig.~\ref{fig:Ising_c} we see the central
charge estimates for the quantum Ising model and the correct
central charge $c=\frac12$ is readily identified for $N=10$ system.
>From these results we conjecture that one may
apply the formula (\ref{Holzhey_formula}) to identify critical points
{\em and} the corresponding central charges by investigating systems
of sizes as low as $N\simeq10$. The main advantage of this is that we can
use an exact numerical diagonalization of the Hamiltonian (possibly
utilizing known information about conservation laws) to obtain critical
information about a larger class of models than what is obtainable
from ordinary means such as fermionizing the $XY$ model reviewed in the
next section.

It is of course well known that global system entanglement increases logarithmically
at critical 
points \cite{Osterloh:2002, Wei04}, and that this scaling
can be utilized \cite{Vidal:2002rm}. But the $\ell$-dependence
reveals much more of the vast information carried by the wave function for a
single system. This seems much more efficient than identifying a slow logarithmic
divergence with system size.

\section{The $\bm{XY}$ and $\bm{XYZ}$ quantum spin chains}

To assure that the results shown in Figs.~\ref{fig:Entanglement}-\ref{fig:Ising_c} 
are more than coincidences, we have studied a larger class of models.
The general translational invariant,
Hermitian spin-$\frac{1}{2}$ chain, with only local and nearest-neighbor interactions,
can (with suitable coordinates) be described by the
Hamiltonian
\begin{equation}
   H = -\sum_{n=0}^{N-1} \left[\sum_{\alpha} f_\alpha\, \sigma^\alpha_n\, \sigma^\alpha_{n+1}
    +\vec{g}\cdot\left(\bm{\sigma}_n \!\bm{\times}\! \bm{\sigma}_{n+1}\right)
    + \bm{h}\cdot\bm{\sigma}_n\right],
   \label{General_H}
\end{equation}
where $\bm{f}$, $\bm{g}$ and $\bm{h}$ are real three-vectors, and
$\bm{\sigma}$ the Pauli spin matrices.
We impose periodic
boundary conditions, $\bm{\sigma}_N \equiv \bm{\sigma_0}$. The
previously studied quantum Ising model corresponds to
the case $f_x=1$, $h_z=\lambda$, with all other parameters zero. This is a special case
of the two-parameter $XY$ quantum spin chain,
defined by $f_x = (1+\gamma)/2$, $f_y = (1-\gamma)/2$, $h_z = \lambda$, i.e.,
\begin{equation}
  H_{XY} = -\sum_{n=0}^{N-1} \left( \frac{1+\gamma}2\,\sigma^x_n\sigma^x_{n+1} +
  \frac{1-\gamma}2\sigma^y_n\, \sigma^y_{n+1} + \lambda\,\sigma^z_n\right),
  \label{H_XY}
\end{equation}
which can be expressed as a quadratic form in fermion variables through
a Jordan-Wigner transformation \cite{footnote3}.

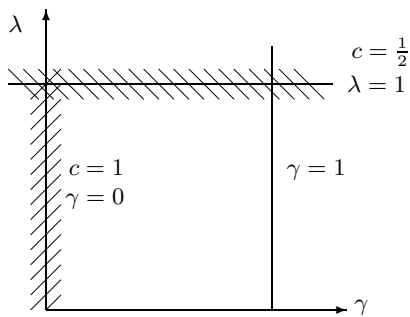
\begin{figure}[h]
  \begin{picture}(50,45)
    \put(5,0){\vector(0,1){40}}
    \put(5,0){\vector(1,0){40}}
    \put(0,30){\line(1,0){43}}
    \put(45,29){\mbox{$\lambda=1$}}
    \put(0,37){\mbox{$\lambda$}}
    \put(46,0){\mbox{$\gamma$}}
    \put(5,0){\line(1,1){2}}
    \multiput(3,0)(0,2){15}{\line(1,1){4}}
    \multiput(0,32)(2,0){20}{\line(1,-1){4}}
    \put(45.6,33.5){\mbox{$c=\frac12$}}
    \put(8,18){\mbox{$c=1$}}
    \put(7.6,14){\mbox{$\gamma=0$}}
    \put(35,0){\line(0,1){35}}
    \put(37,18){\mbox{$\gamma=1$}}
  \end{picture}
  \caption{The critical lines $\lambda=1$ and $\gamma=0$ ($-1\le\lambda\le1$),
    of the $XY$ spin chain are indicated together with their respective
    central charges.
    The second vertical line, $\gamma=1$, corresponds to
    the quantum Ising model.
  }
  \label{fig:Critlines}
\end{figure}

Therefore, many aspects of the $XY$ model are exactly calculable,
making it a good benchmarking tool for our method. 
Symmetry of the space spanned by the anisotropy parameter $\gamma$
and magnetic field $\lambda$ makes it sufficient to consider only
the first quadrant, $\gamma\ge0$, $\lambda\ge0$,
where the model has two critical lines (see Fig.~\ref{fig:Critlines}).
We have investigated how well the critical line $\lambda=1$ is
identified for larger values of $\gamma$. Our findings are that
the behavior at the Ising critical point is not untypical of
the cases considered, but exceptions can be found. One general feature is that
a minimimum in $\varepsilon$ is correlated with a maximum in the
estimated central charge $\hat c$, but that these two points do not
coincide exactly; their difference is a possible measure
of how accurate the critical point has been found. In most cases we
have found the maximum in $\hat c$ to lie closest to the critical
point, but again there are exceptions. There is also a
tendency that the maximum estimated $\hat c$
slightly overshoots its exact value, but since by the Kac formula $c$ is quantized to
$1-6/m(m+1)$ ($m=3,4,\ldots$) in unitary models with
$c<1$ this may be possible to correct. Due to symmetry the method identifies
the $c=1$ critical line $\gamma=0$ exactly;
however with more shallow minima in $\varepsilon$
and a somewhat larger overshoot of $\hat{c}$.

\begin{figure}[t]
   \begin{center}
     \psfrag{A}{$\hat c=0.25$}
     \psfrag{B}{$\hat c=0.5$}
     \psfrag{C}{$\hat c=0.75$}
     \psfrag{D}{$\hat c=1$}
     \psfrag{X}{$\gamma$}
     \psfrag{Y}{$\lambda$}
     \includegraphics[angle=-90]{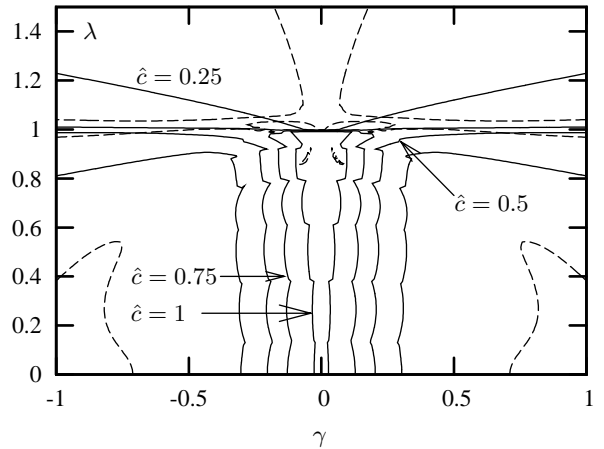}
   \end{center}
   \caption{A contour plot in the parameter space for the $XY$ model, calculated
     for $N=12$. The full lines are contours of the estimated central charge with the 
     indicated values for $\hat c$, while the dashed lines are the contours 
     for the error on the line $\varepsilon=\text{e}^{-20}$. We see a very good 
     correspondence with the known critical lines indicated in Fig.~\ref{fig:Critlines}. 
     (The ``dips'' in the $\hat c$-contour lines
     seen for $\lambda < 1$ are caused by discontinuous changes of the ground state
     \cite{BarouchMccoy71}.)
   }
\label{fig:ContourXY}
\end{figure}

In Fig.~\ref{fig:ContourXY} we show a contour plot of $\hat{c}$ and
$\varepsilon$ over a whole region
of the parameter space in the $XY$ model,
scanned for system size
$N=12$, as would have been realistic in a case where we had no \textit{a priori}
knowledge about the positions of the critical lines (or even their
existence). 
Our conclusion is that such a scan locates the critical lines quite
reliably; the identified regions could next be analyzed more carefully
for larger system sizes. Since there is a crossover between $c=\frac{1}{2}$ and $c=1$ at
$(\gamma,\lambda)=(0,1)$, an interesting question is how
well the method works near this point. We find that the minima in $\varepsilon$ generally
split as we approach this point, probably due to competition between the two critical
lines, thus creating some ambiguity. The region of ambiguity shrinks with increasing
$N$.

By adding another term to (\ref{H_XY}) we obtain the $XYZ$ model,
\begin{equation}
  H_{XYZ}=H_{XY}-\sum_n\Delta\sigma^z_n\sigma^z_{n+1}.
  \label{H_XYZ}
\end{equation}
With $\lambda=0$ the thermodynamics of this model has been found
exactly \cite{Baxter}, Eq.~(\ref{H_XYZ}) then being the Hamiltonian
limit of the transfer matrix of the eight-vertex model. It has a
critical line given by
$\gamma=0$, $\vert\Delta\vert \le \frac{1}{2}$, and a central charge $c=1$.
Also in the isotropic case, $\gamma=0$,
the thermodynamics has been solved exactly.
There is a critical surface given by $\vert\lambda\vert \le 1 -2\vert\Delta\vert$ \cite{footnote4}.

With $\Delta\ne 0$ the Hamiltonian~(\ref{H_XYZ})
can no longer be transformed to a quadratic fermionic form. Thus our entanglement calculations
require a more brute force numerical calculation, starting from a $2^N \times 2^N$ Hamiltonian matrix
(which can be somewhat reduced in size by utilizing all possible
symmetries) \cite{footnote5}.
Thus the required computational resources
grow exponentially with $N$, limiting the current feasible
system size to $N=20$ at best.
\begin{figure}[t]
   \begin{center}
     \psfrag{A}{$\hat c=0.25$}
     \psfrag{B}{$\hat c=0.5$}
     \psfrag{C}{$\hat c=0.75$}
     \psfrag{D}{$\hat c=1$}
     \psfrag{X}{$\gamma$}
     \psfrag{Y}{$\lambda$}
     \includegraphics[angle=-90]{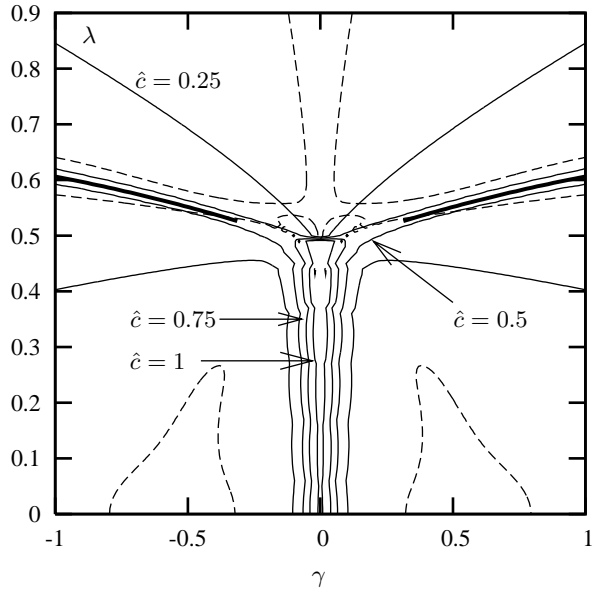}
   \end{center}
   \caption{The same contour plot as in Fig.~\ref{fig:ContourXY}, but now 
for the $XYZ$ model with $\Delta=\frac14$. We see the expected critical 
line at $\gamma=0$ when $\lambda\in[0,1-2\Delta]$ with central charge 1. 
We also see a critical line with central charge $\frac12$ indicated with 
the thick lines. Here the maximum in estimated central charge coincides 
with a minimum in the error.}
\label{fig:ContourXYZ}
\end{figure}

We have analyzed various regions of the three-dimensional parameter space for $N=14$ or less,
and verified that the critical regions mentioned above are
reproduced correctly. In Fig.~\ref{fig:ContourXYZ}
we again show a contour plot of a whole region, scanned for system size
$N=12$ and $\Delta=\frac14$. Here the most interesting feature is
the identification of the $c=\frac{1}{2}$ critical
line, marked with thick lines in Fig.~\ref{fig:ContourXYZ}.
This is a continuation of the $\lambda=1$ critical line of the $XY$ model
to non-zero values of $\Delta$, and has to our knowledge not been calculated before.
In the figure we have not drawn it all the way to the point
$(\gamma,\lambda)=(0,\frac{1}{2})$. This is to indicate the {\em computational
uncertainty\/} which arises near this point, not to indicate that the critical line
ends.

In conclusion, we have found an efficient technique to identify
quantum critical regions of Hamiltonians such as (\ref{General_H}), and others
that show conformal invariance at the critical points. The method
compares the entanglement entropy with the
conformal scaling formula provided by Eq.~(\ref{Holzhey_formula}) of Holzhey
{\em et al.}, and work sufficiently well for small systems to enable us to employ
brute force numerical diagonalization of the Hamiltonian to identify criticality and central
charges in models that are insoluble through other means such as fermionizing.
The method is not limited to spin-$\frac12$ systems or
nearest-neighbor interactions, though it may work better with shorter
range of interactions (since the emergence of conformal symmetry relies
on the existence of an energy-momentum tensor, i.e.,\ on local interactions).

\section*{Acknowledgment}
We thank the Department of Physics, University of Tromsø, where this work was
finalized, for its generous hospitality.

\bibliography{art}

\begin{thebibliography}{17}
\expandafter\ifx\csname natexlab\endcsname\relax\def\natexlab#1{#1}\fi
\expandafter\ifx\csname bibnamefont\endcsname\relax
  \def\bibnamefont#1{#1}\fi
\expandafter\ifx\csname bibfnamefont\endcsname\relax
  \def\bibfnamefont#1{#1}\fi
\expandafter\ifx\csname citenamefont\endcsname\relax
  \def\citenamefont#1{#1}\fi
\expandafter\ifx\csname url\endcsname\relax
  \def\url#1{\texttt{#1}}\fi
\expandafter\ifx\csname urlprefix\endcsname\relax\def\urlprefix{URL }\fi
\providecommand{\bibinfo}[2]{#2}
\providecommand{\eprint}[2][]{\url{#2}}

\bibitem[{\citenamefont{Sachdev}(1999)}]{Sachdev}
\bibinfo{author}{\bibfnamefont{S.}~\bibnamefont{Sachdev}},
  \emph{\bibinfo{title}{Quantum Phase Transitions}}
  (\bibinfo{publisher}{Cambridge University Press},
  \bibinfo{address}{Cambridge, UK}, \bibinfo{year}{1999}).

\bibitem[{\citenamefont{Polyakov}(1970)}]{Polyakov70}
\bibinfo{author}{\bibfnamefont{A.~M.} \bibnamefont{Polyakov}},
  \bibinfo{journal}{JETP Lett.} \textbf{\bibinfo{volume}{12}},
  \bibinfo{pages}{381} (\bibinfo{year}{1970}).

\bibitem[{\citenamefont{Belavin
  et~al.}(1984{\natexlab{a}})\citenamefont{Belavin, Polyakov, and
  Zamolodchikov}}]{BPZ84a}
\bibinfo{author}{\bibfnamefont{A.~A.} \bibnamefont{Belavin}},
  \bibinfo{author}{\bibfnamefont{A.~M.} \bibnamefont{Polyakov}},
  \bibnamefont{and} \bibinfo{author}{\bibfnamefont{A.~B.}
  \bibnamefont{Zamolodchikov}}, \bibinfo{journal}{J. Stat Phys}
  \textbf{\bibinfo{volume}{34}}, \bibinfo{pages}{763}
  (\bibinfo{year}{1984}{\natexlab{a}}).

\bibitem[{\citenamefont{Belavin
  et~al.}(1984{\natexlab{b}})\citenamefont{Belavin, Polyakov, and
  Zamolodchikov}}]{BPZ84b}
\bibinfo{author}{\bibfnamefont{A.~A.} \bibnamefont{Belavin}},
  \bibinfo{author}{\bibfnamefont{A.~M.} \bibnamefont{Polyakov}},
  \bibnamefont{and} \bibinfo{author}{\bibfnamefont{A.~B.}
  \bibnamefont{Zamolodchikov}}, \bibinfo{journal}{Nucl. Phys. B}
  \textbf{\bibinfo{volume}{241}}, \bibinfo{pages}{333}
  (\bibinfo{year}{1984}{\natexlab{b}}).

\bibitem[{\citenamefont{Nielsen and Chuang}(2000)}]{Nielsen&Chuang}
\bibinfo{author}{\bibfnamefont{M.~A.} \bibnamefont{Nielsen}} \bibnamefont{and}
  \bibinfo{author}{\bibfnamefont{I.~L.} \bibnamefont{Chuang}},
  \emph{\bibinfo{title}{Quantum Computation and Quantum Information}}
  (\bibinfo{publisher}{Cambridge University Press},
  \bibinfo{address}{Cambridge, UK}, \bibinfo{year}{2000}).

\bibitem[{\citenamefont{Holzhey et~al.}(1994)\citenamefont{Holzhey, Larsen, and
  Wilczek}}]{Holzhey94}
\bibinfo{author}{\bibfnamefont{C.}~\bibnamefont{Holzhey}},
  \bibinfo{author}{\bibfnamefont{F.}~\bibnamefont{Larsen}}, \bibnamefont{and}
  \bibinfo{author}{\bibfnamefont{F.}~\bibnamefont{Wilczek}},
  \bibinfo{journal}{Nucl.Phys. B} \textbf{\bibinfo{volume}{424}},
  \bibinfo{pages}{443} (\bibinfo{year}{1994}).

\bibitem[{\citenamefont{Calabrese and Cardy}(2004)}]{Calabrese04}
\bibinfo{author}{\bibfnamefont{P.}~\bibnamefont{Calabrese}} \bibnamefont{and}
  \bibinfo{author}{\bibfnamefont{J.}~\bibnamefont{Cardy}},
  \bibinfo{journal}{Journal of Statistical Mechanics: Theory and Experiment} p.
  \bibinfo{pages}{P06002} (\bibinfo{year}{2004}), \eprint{hep-th/0405152}.

\bibitem[{foo({\natexlab{a}})}]{footnote1}
\bibinfo{note}{By {\em critical parameters\/} we mean parameter values
  corresponding to critical points in the thermodynamic limit.}

\bibitem[{foo({\natexlab{b}})}]{footnote2}
\bibinfo{note}{Consider transfer matrices for the 2D Ising model on a square
  lattice, with bond strengths $K$ and $L$ in the two directions, and define
  $\lambda=(\sinh 2K \sinh 2L)^{-1}$. All transfer matrices with the same value
  of $\lambda$ commute, and thus have the same eigenvectors. This is because
  they may be written in the form $\exp(\tau H_{\mathrm{Ising}})$ for some
  $\tau$ (modulo a constant prefactor).}

\bibitem[{\citenamefont{Osterloh et~al.}(2002)\citenamefont{Osterloh, Amico,
  Falci, and Fazio}}]{Osterloh:2002}
\bibinfo{author}{\bibfnamefont{A.}~\bibnamefont{Osterloh}},
  \bibinfo{author}{\bibfnamefont{L.}~\bibnamefont{Amico}},
  \bibinfo{author}{\bibfnamefont{G.}~\bibnamefont{Falci}}, \bibnamefont{and}
  \bibinfo{author}{\bibfnamefont{R.}~\bibnamefont{Fazio}},
  \bibinfo{journal}{Nature} \textbf{\bibinfo{volume}{416}},
  \bibinfo{pages}{608} (\bibinfo{year}{2002}).

\bibitem[{\citenamefont{Wei et~al.}(2004)\citenamefont{Wei, Das, Mukhopadyay,
  Vishveshwara, and Goldbart}}]{Wei04}
\bibinfo{author}{\bibfnamefont{T.-C.} \bibnamefont{Wei}},
  \bibinfo{author}{\bibfnamefont{D.}~\bibnamefont{Das}},
  \bibinfo{author}{\bibfnamefont{S.}~\bibnamefont{Mukhopadyay}},
  \bibinfo{author}{\bibfnamefont{S.}~\bibnamefont{Vishveshwara}},
  \bibnamefont{and} \bibinfo{author}{\bibfnamefont{P.~M.}
  \bibnamefont{Goldbart}} (\bibinfo{year}{2004}), \eprint{quant-ph/0405162}.

\bibitem[{\citenamefont{Vidal et~al.}(2002)\citenamefont{Vidal, Latorre, Rico,
  and Kitaev}}]{Vidal:2002rm}
\bibinfo{author}{\bibfnamefont{G.}~\bibnamefont{Vidal}},
  \bibinfo{author}{\bibfnamefont{J.~I.} \bibnamefont{Latorre}},
  \bibinfo{author}{\bibfnamefont{E.}~\bibnamefont{Rico}}, \bibnamefont{and}
  \bibinfo{author}{\bibfnamefont{A.}~\bibnamefont{Kitaev}},
  \bibinfo{journal}{Phys. Rev. Lett.} \textbf{\bibinfo{volume}{90}},
  \bibinfo{pages}{227902} (\bibinfo{year}{2002}).

\bibitem[{foo({\natexlab{c}})}]{footnote3}
\bibinfo{note}{With a periodic chain this must be carried out separately in
  each eigenspace of the operator $\prod_{n=0}^{N-1} \sigma_n^z$.}

\bibitem[{\citenamefont{Barouch and McCoy}(1971)}]{BarouchMccoy71}
\bibinfo{author}{\bibfnamefont{E.}~\bibnamefont{Barouch}} \bibnamefont{and}
  \bibinfo{author}{\bibfnamefont{B.~M.} \bibnamefont{McCoy}},
  \bibinfo{journal}{Phys. Rev. A} \textbf{\bibinfo{volume}{3}},
  \bibinfo{pages}{786} (\bibinfo{year}{1971}).

\bibitem[{\citenamefont{Baxter}(1982)}]{Baxter}
\bibinfo{author}{\bibfnamefont{R.~J.} \bibnamefont{Baxter}},
  \emph{\bibinfo{title}{Exactly Solved Models in Statistical Mechanics}}
  (\bibinfo{publisher}{Academic Press}, \bibinfo{address}{New York},
  \bibinfo{year}{1982}).

\bibitem[{foo({\natexlab{d}})}]{footnote4}
\bibinfo{note}{There are of course many more critical regions in the
  three-dimensional parameter space, related to those mentioned by permutation
  and other symmetries.}

\bibitem[{foo({\natexlab{e}})}]{footnote5}
\bibinfo{note}{All our numerical computations are done on ordinary PC's using
  Lapack diagonalization routines.}

\end{thebibliography}

\end{document}